\documentclass[12pt]{article}
\input epsf.tex
\begin{document}
\newcommand{\beq}{\begin{equation}}
\newcommand{\eeq}{\end{equation}}
\newcommand{\beqa}{\begin{eqnarray}}
\newcommand{\eeqa}{\end{eqnarray}}
\newcommand{\beqar}{\begin{eqnarray*}}
\newcommand{\eeqar}{\end{eqnarray*}}
\newcommand{\al}{\alpha}
\newcommand{\be}{\beta}
\newcommand{\del}{\delta}
\newcommand{\D}{\Delta}
\newcommand{\eps}{\epsilon}
\newcommand{\ga}{\gamma}
\newcommand{\Ga}{\Gamma}
\newcommand{\ka}{\kappa}
\newcommand{\inn}{\!\cdot\!}
\newcommand{\h}{\eta}
\newcommand{\kk}{\varphi}
\newcommand\F{{}_3F_2}
\newcommand{\la}{\lambda}
\newcommand{\La}{\Lambda}
\newcommand{\na}{\nabla}
\newcommand{\Om}{\Omega}
\newcommand{\p}{\phi}
\newcommand{\sig}{\sigma}
\renewcommand{\t}{\theta}
\newcommand{\z}{\zeta}
\newcommand{\ssc}{\scriptscriptstyle}
\newcommand{\eg}{{\it e.g.,}\ }
\newcommand{\ie}{{\it i.e.,}\ }
\newcommand{\labell}[1]{\label{#1}} 
\newcommand{\reef}[1]{(\ref{#1})}
\newcommand{\labels}[1]{\vskip-2ex$_{#1}$\label{#1}} 
\newcommand\prt{\partial}
\newcommand\veps{\varepsilon}
\newcommand\ls{\ell_s}
\newcommand\cF{{\cal F}}
\newcommand\cA{{\cal A}}
\newcommand\cS{{\cal S}}
\newcommand\cH{{\cal H}}
\newcommand\cC{{\cal C}}
\newcommand\cL{{\cal L}}
\newcommand\cG{{\cal G}}
\newcommand\cI{{\cal I}}
\newcommand\cl{{\iota}}
\newcommand\cP{{\cal P}}
\newcommand\cV{{\cal V}}
\newcommand\cg{{\it g}}
\newcommand\cR{{\cal R}}
\newcommand\cB{{\cal B}}
\newcommand\cO{{\cal O}}
\newcommand\tcO{{\tilde {{\cal O}}}}
\newcommand\bz{\bar{z}}
\newcommand\bw{\bar{w}}
\newcommand\hF{\hat{F}}
\newcommand\hA{\hat{A}}
\newcommand\hT{\hat{T}}
\newcommand\htau{\hat{\tau}}
\newcommand\hD{\hat{D}}
\newcommand\hf{\hat{f}}
\newcommand\hg{\hat{g}}
\newcommand\hp{\hat{\phi}}
\newcommand\hh{\hat{h}}
\newcommand\ha{\hat{a}}
\newcommand\hQ{\hat{Q}}
\newcommand\hP{\hat{\Phi}}
\newcommand\hb{\hat{b}}
\newcommand\hc{\hat{c}}
\newcommand\hd{\hat{d}}
\newcommand\hS{\hat{S}}
\newcommand\hX{\hat{X}}
\newcommand\tL{\tilde{\cal L}}
\newcommand\hL{\hat{\cal L}}
\newcommand\tG{{\widetilde G}}
\newcommand\tg{{\widetilde g}}
\newcommand\tphi{{\widetilde \phi}}
\newcommand\tPhi{{\widetilde \Phi}}
\newcommand\te{{\tilde e}}
\newcommand\tk{{\tilde k}}
\newcommand\tf{{\tilde f}}
\newcommand\tF{{\widetilde F}}
\newcommand\tK{{\widetilde K}}
\newcommand\tE{{\widetilde E}}
\newcommand\tpsi{{\tilde \psi}}
\newcommand\tX{{\widetilde X}}
\newcommand\tD{{\widetilde D}}
\newcommand\tO{{\widetilde O}}
\newcommand\tS{{\tilde S}}
\newcommand\tB{{\widetilde B}}
\newcommand\tA{{\widetilde A}}
\newcommand\tT{{\widetilde T}}
\newcommand\tC{{\widetilde C}}
\newcommand\tV{{\widetilde V}}
\newcommand\thF{{\widetilde {\hat {F}}}}
\newcommand\Tr{{\rm Tr}}
\newcommand\tr{{\rm tr}}
\newcommand\STr{{\rm STr}}
\newcommand\M[2]{M^{#1}{}_{#2}}
\parskip 0.3cm

\vspace*{1cm}

\begin{center}
{\bf \Large   T-duality of the Riemann curvature \\corrections to  supergravity     }

\vspace*{1cm}

{  Mohammad R. Garousi\footnote{garousi@ferdowsi.um.ac.ir} }\\
\vspace*{1cm}
{ Department of Physics, Ferdowsi University of Mashhad,\\ P.O. Box 1436, Mashhad, Iran}
\\
\vspace{2cm}

\end{center}

\begin{abstract}
\baselineskip=18pt

We examine    the sigma model Riemann curvature  corrections to the supergravity action    under  
 T-duality transformations. Using the  compatibility of the effective action with on-shell linear T-duality and with the S-matrix calculations  as  guiding principles, we have incorporated  in this action  the couplings of four $B$-field strengths and the couplings of two Riemann curvatures and two $B$-field strengths   at order $\alpha'^3$. Using the S-matrix calculations we have also found new dilaton couplings in the string frame at this order. 
 

\end{abstract}
Keywords: T-duality, S-matrix 

\setcounter{page}{0}
\setcounter{footnote}{0}
\newpage
\section{Introduction and results} \label{intro}

A standard method in string theory for finding the higher derivative corrections to the supergravity action \cite{Schwarz:1983qr,Campbell:1984zc} is the scattering amplitude calculation \cite{Green:1981xx,Green:1981ya}.  The $\alpha'^3$ corrections to the Einstein-Hilbert action have been found in \cite{ Gross:1986iv} by analyzing the sphere-level four-graviton scattering amplitude in type II superstring theory. The result in the eight-dimensional transverse space of the light-cone formalism, is a polynomial in the Riemann curvature tensors 
\beqa
Y&\sim& t^{i_1\cdots i_8}t_{j_1\cdots j_8}\cR_{i_1i_2}{}^{j_1j_2}\cdots \cR_{i_7i_8}{}^{j_7j_8}\labell{Y1}
\eeqa
where $t^{i_1\cdots i_8}$ is a tensor in eight dimensions which includes the eight-dimensional Levi-Civita tensor \cite{Gross:1986iv}. This $SO(8)$ invariant  Lagrangian has been extended to Lorentz invariant form  in \cite{Grisaru:1986vi,Freeman:1986zh}
\beqa
\cL&=& \frac{\gamma e^{-2\phi_0}}{\kappa^2} [\cR_{hmnk}\cR_p{}^{mn}{}_q\cR^{hrsp}\cR^q{}_{rs}{}^k+\frac{1}{2}\cR_{hkmn}\cR_{pq}{}^{mn}\cR^{hrsp}\cR^q{}_{rs}{}^k+\cdots]\labell{Y2}
\eeqa
where $\gamma=\frac{1}{8}\alpha'^3\z(3)$, $e^{-2\phi_0}$ is the dilaton background corresponding to the sphere-level scattering amplitude,  and dots represent terms containing the Ricci and scalar curvature tensors. These terms can not be captured by the four-graviton scattering amplitude as they are zero on-shell. They can be absorbed by the Einstein-Hilbert action in field redefinition of the metric $G\rightarrow G+\delta G$ which does not alter the scattering calculation \cite{ Gross:1986iv}. The above $R^4$ couplings reproduce the sigma-model beta function \cite{Grisaru:1986vi,Freeman:1986zh}.

Unlike the Einstein-Hilbert Lagrangian, there are  different Lorentz invariant expressions for the Riemann curvature couplings  at order $\alpha'^3$ \cite{Myers:1987qx,Gross:1986mw}. They are  related to \reef{Y2} via some identities  involving the Riemann curvature tensors and  some  couplings involving the Ricci and scalar curvature tensors \cite{Myers:1987qx}. The  Ricci and scalar curvature couplings  can be eliminated by field redefinitions, however, the identities  involving the Riemann curvature tensors hold only at four graviton levels \cite{Myers:1987qx}. As a result, there may be some other four Riemann curvature couplings  in \reef{Y2}  which can be found  by studying five-graviton scattering amplitude in which we are not interested in this paper. 

 For the  Lagrangian presented in \cite{Myers:1987qx,Gross:1986mw}, a proposal has been given in \cite{Gross:1986mw} for including the $B$-field and the dilaton   into  the action  which is a prescription for generalizing   the Riemann curvature tensor to include the first derivative of the $B$-field strength and the second derivative of the dilaton. While this prescription gives the correct $B$-field couplings for the Lagrangian  given in \cite{Myers:1987qx,Gross:1986mw}, we will show that it does not work for the   Lagrangian \reef{Y2}.  In this paper we would like to extend this Lagrangian   to include the B-field and dilaton by using the compatibility of the couplings \reef{Y2} with the T-duality \cite{Kikkawa:1984cp,TB,Giveon:1994fu,Alvarez:1994dn}  and by using the scattering amplitude calculations \cite{Schwarz:1982jn,Gross:1986iv}. Similar calculations have been done in \cite{Garousi:2009dj,Garousi:2010ki,Becker:2010ij,Garousi:2010rn,Garousi:2010bm,Garousi:2011ut,Becker:2011bw,Becker:2011ar,Velni:2012sv,McOrist:2012yc} to extend the curvature couplings on the world volume of D-brane to all other massless fields.

The outline of the paper is as follows: We begin in section 2 by reviewing the T-duality transformations and finding the transformation of the linearized curvature tensors under linear T-duality. In section 3, we review the sphere-level scattering amplitude of four massless NS-NS states in type II superstring theory and reconfirm that this amplitude at order $\alpha'^3$ produces the couplings \reef{Y2}. In section 4, we reduce the 10-dimensional couplings \reef{Y2} to 9 dimensions to find the $\cR_y\cR_y\cR_y\cR_y$ couplings where $\cR_y$ is the   Riemann tensor with one Killing index. The consistency of these couplings with the linear T-duality is used to find the following couplings:
\beqa
\cL&\!\!\!\!\supset\!\!\!\!&\frac{\gamma  e^{-2\phi_0}}{16\kappa^2} \bigg[- \cH_{h p r;k} \cH_n {}^{p r}{}_{;q} \cH^{k m s;q} \cH_{m}{}^ n {}_s{}^{;h}- \cH_{h m n;k} \cH^{h n p;q} \cH^{k}{}_{ p s;r} \cH^{m}{}_ q {}^{s;r}\nonumber\\
&&\qquad\qquad+ \cH_{k m n;h} \cH^{h p q;n} \cH^{k}{}_ {p s;r} \cH^{m}{}_ q {}^{s;r}\bigg]\labell{H40}
\eeqa
where $\cH$ is the B-field strength, $\cH_{abc}=B_{ab,c}+B_{ca,b}+B_{bc,a}$. As usual, the commas and the  semicolons represent partial and covariant derivatives, respectively. We have also explicitly confirmed the above couplings with the S-matrix element of four $B$-fields in type II superstring theory. In section 5, we consider the consistency of the couplings $\cR\cR\cR_y\cR_y$ and $\cH\cH\cH_y\cH_y$   with the linear T-duality to find $\cR\cR\cH\cH$ couplings. The couplings $\cH\cH \cR_y\cR_y$ and $\cR\cR\cH_y\cH_y$ must be   the T-duality transformations of $\cH\cH\cH_y\cH_y$ and $\cR\cR\cR_y\cR_y$, respectively. Moreover, the couplings $\cH_y\cH_y \cR_y\cR_y$ and $\cH \cH_y \cR \cR_y$ each must be invariant under the T-duality transformations. Imposing these conditions and using on-shell relations, we have found the following couplings:
\beqa
\cL&\!\!\!\!\supset\!\!\!\!&\frac{\gamma  e^{-2\phi_0}}{2\kappa^2}\bigg[ \cR_{ h mk n } \cR^{ m pn q } \cH^{k r s}{}_{;q} \cH_{p r s}{}^{;h}-2 \cR_{ h rp s } \cR^{ q rk s } \cH^{h}{}_{ k n;m} \cH^{n p}{}_ q{}^{;m}\nonumber\\
&&\qquad\quad+2 \cR_{ m pn q } \cR^{ q rk s } \cH^{h}{}_ k{}^{ n;m} \cH_{h}{}^ p{}_{ s;r}+ \cR_{ m np q } \cR^{ q rk s } \cH^{h m n}{}_{;k} \cH_{h}{}^ p{}_{ s;r}\nonumber\\
&&\qquad\quad+2 \cR_{ m pn q } \cR^{ q rk s } \cH_{k}{}^{ m n;h} \cH^{p}{}_{ r s;h}+ \cR_{ h mk n } \cR^{ m pn q } \cH^{h}{}_{ p s;r} \cH^{k}{}_q {}^{s;r} \nonumber\\
&&\qquad\quad-2 \cR_{ m pn q } \cR^{ q rk s } \cH_{k}{}^{ m n;h} \cH_{h }{}^p{}_{ s;r}+2 \cR_{ h mk n } \cR^{ m pn q } \cH^{k}{}_{ q s;r} \cH_{p}{}^{ r s;h}\nonumber\\
&&\qquad\quad- 6 \cR_{ h rp s } \cR^{ q rk s } \cH^{h}{}_{ k n;m} \cH^{m n}{}_ q{}^{;p}\bigg]\labell{rrhh0}
\eeqa
In section 6, we discuss the dilaton couplings. We argue that many terms of the dilaton amplitudes are reproduced by transforming the string frame couplings \reef{Y2} and \reef{rrhh0} to the Einstein frame. However, there are some terms in the scattering amplitudes that cannot be reproduced in this way. The scattering amplitude of two dilatons and two gravitons produces the following couplings as well as the couplings in \reef{Y2}:
\beqa
\cL&\!\!\!\!\supset\!\!\!\!&-\frac{\gamma  e^{-2\phi_0}}{16\kappa^2} \bigg[\cR^{h k }{}_{m n} \cR^{m n p q} \Phi_{;hp} \Phi_{;kq} \nonumber\\
&& \qquad\qquad+2 \cR^{h}{}_ m {}^k{}_{ n} \cR^{m p n q} \Phi_{;hp}\Phi_{;kq} +2 \cR^{h}{}_ m {}^k{}_{ n} \cR^{q m p n} \Phi_{;hp}\Phi_{;kq} \bigg]\labell{delRR0}
\eeqa
The scattering amplitude of four dilatons produces the following   couplings:
\beqa
\cL&\!\!\!\!\supset\!\!\!\!&-\frac{\gamma  e^{-2\phi_0}}{ 64\kappa^2}\bigg[ \Phi_{;hn}  \Phi^{;hs} \Phi^{;nq}  \Phi_{;qs} - \Phi_{;mn} \Phi^{;mn} \Phi_{;rs} \Phi^{;rs}  \bigg]\labell{del4}
\eeqa
And the scattering amplitude of two dilatons and two B-fields produces the couplings:
\beqa
\cL&\!\!\!\!\supset\!\!\!\!&\frac{\gamma  e^{-2\phi_0}}{ 16\kappa^2}\bigg[6 \Phi_{;hp}  \Phi_{;kq} \cH^{h{  }k n;m} \cH_{m n}{}^{ q;p}+2\Phi_{;hp}   \Phi_{;kq}  \cH^{h{  }k n;m} \cH_{n}{}^{ p q}{}_{;m}\labell{phiH}- \Phi_{;hk}  \Phi_{;pq} \cH^{h{  }p n;m} \cH^{k q}{}_{ n;m}\nonumber\\&&\qquad\qquad-2 \Phi_{;hk}  \Phi_{;pq}  \cH^{k q n;m} \cH^{p}{}_{ m n}{}^{;h } - \Phi_{;hk}  \Phi_{;pq}  \cH^{k m n;q} \cH^{p}{}_{ m n}{}^{;h }\bigg]\nonumber
\eeqa
In section 7, we briefly discuss our results.

\section{T-duality}

The full set of nonlinear T-duality transformations   have been found in \cite{TB}.   When  the T-duality transformation acts along the Killing coordinate $y$,  the massless NS-NS fields   transform as:
\beqa
e^{2\tPhi}=\frac{e^{2\Phi}}{G_{yy}}&;& 
\tG_{yy}=\frac{1}{G_{yy}}\nonumber\\
\tG_{\mu y}=\frac{B_{\mu y}}{G_{yy}}&;&
\tG_{\mu\nu}=G_{\mu\nu}-\frac{G_{\mu y}G_{\nu y}-B_{\mu y}B_{\nu y}}{G_{yy}}\nonumber\\
\tB_{\mu y}=\frac{G_{\mu y}}{G_{yy}}&;&
\tB_{\mu\nu}=B_{\mu\nu}-\frac{B_{\mu y}G_{\nu y}-G_{\mu y}B_{\nu y}}{G_{yy}}\labell{nonlinear}
\eeqa
where $\mu,\nu$ denote any coordinate directions other than $y$. In above transformation the metric is given in the string frame. If $y$ is identified on a circle of radius $\rho$, \ie $y\sim y+2\pi \rho$, then after T-duality the radius becomes $\tilde{\rho}=\alpha'/\rho$. The string coupling $g=e^{\phi_0}$ is also shifted as $\tilde{g}=g\sqrt{\alpha'}/\rho$.

We would like to study the T-dual Ward identity \cite{ Garousi:2011we,Velni:2012sv} of  the scattering amplitude of four gravitons, so we need the above transformations at the linear order. Assuming that the NS-NS  fields are small perturbations around the background, \ie
\beqa
G_{\mu\nu}&=&\eta_{\mu\nu}+2\kappa h_{\mu\nu}\,;\,\,G_{yy}\,=\,\frac{\rho^2}{\alpha'}(1+2\kappa h_{yy})\,;\,\,G_{\mu y}\,=\,2\kappa h_{\mu y}\nonumber\\
B_{\mu\nu}&=&2\kappa b_{\mu\nu}\,;\,B_{\mu y}\,=\,2\kappa b_{\mu y}\nonumber\\
\Phi&=&\phi_0+\sqrt{2}\kappa \phi
\eeqa
the   transformations \reef{nonlinear} take the following linear form for the perturbations:
\beqa
&&
 \sqrt{2}\tilde{\phi}=\sqrt{2}\phi - h_{yy},\,\tilde{h}_{yy}=-h_{yy},\, \tilde{h}_{\mu y}=b_{\mu y},\, \tilde{b}_{\mu y}=h_{\mu y},\,\tilde{h}_{\mu\nu}=h_{\mu\nu},\,\tilde{b}_{\mu\nu}=b_{\mu\nu}\labell{linear}
\eeqa

 To study the linear T-duality  of the couplings \reef{Y2}, it is convenient   to   find the T-duality transformation  of the linearized  curvature tensors. The Riemann tensor at the linear order in graviton is given by
\beqa
R_{abcd}&=& \kappa(h_{ad,bc}+h_{bc,ad}-h_{ac,bd}-h_{bd,ac})
\eeqa
  In the case that one of its indices is the $y$-index, \ie $R_{abc y}$ where $y$ is the Killing direction along which the T-duality is to be performed, it becomes
\beqa
R_{abc y}&=& \kappa(h_{a y,bc}+h_{bc,a y}-h_{ac,b y}-h_{b y,ac})= \kappa(h_{a y,bc}-h_{b y,ac})
\eeqa
where the second equality assumes that all fields are independent of the T-dual coordinate $y$. This becomes after T-duality,
\beqa
  \kappa (b_{a y,bc}-b_{b y,ac})
\eeqa
to which one may trivially add $ \kappa  b_{ba,yc}$ since the fields are assumed independent of $y$,  and hence complete the exterior derivative. Therefore, the linearized transformation of the Riemann tensor with one $y$-index is 
\beqa
R_{abc y}\leftrightarrow  - \kappa H_{ab y,c}\labell{TR}
\eeqa
where $H_{\mu\nu\rho}= b_{\mu\nu,\rho}+b_{\rho\mu,\nu}+b_{\nu\rho,\mu}$. Here the arrow goes in both directions since the derivation can clearly be run in reverse and hence these two expressions are exchanged under T-duality. 

In the case that two indices of the Riemann tensor are the $y$ indices, \ie $R_{ayby}$, it becomes
\beqa
R_{ayby}=-\frac{\kappa \rho^2}{\alpha'}h_{yy,ab}\leftrightarrow \frac{\kappa \rho^2}{\alpha'} h_{yy,ab}=-R_{ayby}\labell{TR1}
\eeqa
where again derivatives of $h$ with respect to $y$ have been dropped and added in order to complete the curvatures. Note that due to the complete antisymmetric of the B-field strength no analogous terms with a double $y$ index will be relevant. That is $H_{ayy}=0$ by the antisymmetry of the indices. Similarly, $R_{abyy}=0 $ by antisymmetry. 
 
The transformation of the Ricci curvature tensor involves the dilaton as well as the B-field strength. To see this consider the case that none of the indices of the Ricci tensor carries the $y$ index. It transforms as
\beqa
R_{ab}=\eta^{cd}R_{cadb}+\frac{\alpha'}{\rho^2}R_{yayb}=\eta^{cd}R_{cadb}-\kappa h_{yy,ab}\rightarrow \eta^{cd}R_{cadb}+\kappa h_{yy,ab}=R_{ab}+2\kappa h_{yy,ab}
\eeqa
where in the first identity we have separated the contracted indices along and orthogonal to  $y$. The last term is not tensor, so there must be another term whose T-duality cancels that term. Using the linear transformation of the dilaton \reef{linear}, one finds the following combination is invariant:
\beqa
R_{ab}+2\sqrt{2}\kappa \phi_{,ab}\leftrightarrow R_{ab}+2\sqrt{2}\kappa \phi_{,ab}\labell{TR2}
\eeqa
Similarly the transformation of the Ricci curvature when it carries one or two $y$ indices, and the transformation of the scalar curvature are
\beqa
R_{ay}&\leftrightarrow &\kappa H_{aby}{}^{,b}\nonumber\\
R_{yy}&\leftrightarrow& -R_{yy}\nonumber\\
R+4\sqrt{2}\kappa \phi_{,a}{}^{a}&\leftrightarrow &R+4\sqrt{2}\kappa \phi_{,a}{}^{a}\labell{TR3}
\eeqa
The last transformation in particular indicates that the supergravity Lagrangian must include a Laplacian of the dilaton to be invariant under the T-duality (see equation (1.10) in \cite{Hohm:2010jy} for the presence of such term in the T-dual Lagrangian at leading order of $\alpha'$.). 

To extend a coupling to a set of couplings which   
  are invariant under linear T-duality,    we first use the dimensional reduction to reduce the 10-dimensional couplings to 9-dimensional couplings, \ie separate the  indices  along and orthogonal to $y$,     and then apply the  above T-duality transformations.  If the original coupling is not invariant under the T-duality, one must add new terms to make them invariant.  

\section{Four-point amplitude}

The scattering amplitude of four massless NS-NS states with polarization tensors $\veps^{ab}$ in covariant formalism is given by \cite{Schwarz:1982jn}
\beqa
{\cal A} =-\frac{\alpha'^3\kappa^2e^{-2\phi_0}}{16} \frac{\Gamma(-s/8)\Gamma(-t/8)\Gamma(-u/8)}{\Gamma(1+s/8)\Gamma(1+t/8)\Gamma(1+u/8)}\veps_1^{a_1b_1}\veps_2^{a_2b_2}\veps_3^{a_3b_3}\veps_4^{a_4b_4}K_{a_1a_2a_3a_4}K_{b_1b_2b_3b_4}\labell{amp1}
\eeqa 
There is also a factor of delta function $\delta^{10}(k_1+k_2+k_3+k_4)$ imposing conservation of momentum. The Mandelstam variables $s=-4\alpha'k_1\inn k_2$, $t=-4\alpha'k_1\inn k_3$, $u=-4\alpha'k_2\inn k_3$ satisfy $s+t+u=0$, and 
\beqa
K_{a_1a_2a_3a_4}&\!\!\!\!\!=\!\!\!\!\!&4\bigg[- k_2.k_1 k_3.k_1 \eta _{{a1a4}} \eta _{{a2a3}}- k_2.k_1 k_2.k_3 \eta _{{a1a3}} \eta _{{a2a4}}- k_2.k_3 k_3.k_1 \eta _{{a1a2}} \eta _{{a3a4}}\nonumber\\&&+ k_2.k_1 \eta _{{a1a4}} \left(k_1\right)_{{a2}} \left(k_1\right)_{{a3}}+ k_3.k_1 \eta _{{a1a4}} \left(k_1\right)_{{a2}} \left(k_1\right)_{{a3}}- k_2.k_1 \eta _{{a2a4}} \left(k_1\right)_{{a3}} \left(k_2\right)_{{a1}}\nonumber\\&&- k_3.k_1 \eta _{{a2a4}} \left(k_1\right)_{{a3}} \left(k_2\right)_{{a1}}+ k_3.k_1 \eta _{{a1a4}} \left(k_1\right)_{{a2}} \left(k_2\right)_{{a3}}- k_3.k_1 \eta _{{a2a4}} \left(k_2\right)_{{a1}} \left(k_2\right)_{{a3}}\nonumber\\&&+ k_2.k_1 \eta _{{a1a2}} \left(k_1\right)_{{a3}} \left(k_2\right)_{{a4}}+ k_3.k_1 \eta _{{a1a2}} \left(k_1\right)_{{a3}} \left(k_2\right)_{{a4}}+ k_3.k_1 \eta _{{a1a2}} \left(k_2\right)_{{a3}} \left(k_2\right)_{{a4}}\nonumber\\&&- k_2.k_1 \eta _{{a3a4}} \left(k_1\right)_{{a2}} \left(k_3\right)_{{a1}}- k_3.k_1 \eta _{{a3a4}} \left(k_1\right)_{{a2}} \left(k_3\right)_{{a1}}+ k_2.k_1 \eta _{{a2a4}} \left(k_2\right)_{{a3}} \left(k_3\right)_{{a1}}\nonumber\\&&- k_2.k_1 \eta _{{a2a3}} \left(k_2\right)_{{a4}} \left(k_3\right)_{{a1}}+ k_2.k_1 \eta _{{a1a4}} \left(k_1\right)_{{a3}} \left(k_3\right)_{{a2}}+ k_3.k_1 \eta _{{a3a4}} \left(k_2\right)_{{a1}} \left(k_3\right)_{{a2}}\nonumber\\&&+ k_2.k_1 \eta _{{a1a3}} \left(k_2\right)_{{a4}} \left(k_3\right)_{{a2}}- k_2.k_1 \eta _{{a3a4}} \left(k_3\right)_{{a1}} \left(k_3\right)_{{a2}}+ k_2.k_1 \eta _{{a1a3}} \left(k_1\right)_{{a2}} \left(k_3\right)_{{a4}}\nonumber\\&&+ k_3.k_1 \eta _{{a1a3}} \left(k_1\right)_{{a2}} \left(k_3\right)_{{a4}}- k_3.k_1 \eta _{{a2a3}} \left(k_2\right)_{{a1}} \left(k_3\right)_{{a4}}+ k_3.k_1 \eta _{{a1a2}} \left(k_2\right)_{{a3}} \left(k_3\right)_{{a4}}\nonumber\\&&+ k_2.k_1 \eta _{{a1a3}} \left(k_3\right)_{{a2}} \left(k_3\right)_{{a4}}\bigg]\labell{kin1}
\eeqa
The on-shell conditions are $k_i^2=k_i\inn\veps_i=\veps_i\inn k_i=0$. The polarization tensor is symmetric and traceless for graviton, antisymmetric for B-field and for dilaton it is 
\beqa
\veps^{ab}= \frac{\phi}{\sqrt{8}}(\eta^{ab}-k^a\ell^b-k^b\ell^a)\labell{del}
\eeqa
where $\ell^a$ is an auxiliary vector which satisfies $k\inn\ell=1$ and $\phi$ is the dilaton polarization which is one. In  equation \reef{kin1} we have used the conservation of momentum   to write the amplitude in terms of momentum $k_1,k_2,k_3$. We have also used the on-shell conditions to rewrite $k_1\inn\veps_4=-k_2\inn\veps_4-k_3\inn\veps_4$, similarly for $\veps_4\cdot k_1$. We have normalized the amplitude \reef{amp1} to be consistent with the normalization factor in the  couplings \reef{Y2}.

The coupling \reef{Y1} has been found in \cite{Gross:1986iv}  from  the amplitude \reef{amp1} by expanding  the gamma functions at low energy  
\beqa
\frac{\Gamma(-s/8)\Gamma(-t/8)\Gamma(-u/8)}{\Gamma(1+s/8)\Gamma(1+t/8)\Gamma(1+u/8)}&=&-\frac{2^9}{stu}-2\z(3)+\cdots
\eeqa
The first term corresponds to the massless poles in the four-point function which are reproduced by the Einstein-Hilbert action \cite{Sannan:1986tz}, and the second term,
\beqa
\Delta{\cal A}&=& \gamma \kappa^2e^{-2\phi_0} \veps_1^{a_1b_1}\veps_2^{a_2b_2}\veps_3^{a_3b_3}\veps_4^{a_4b_4}K_{a_1a_2a_3a_4}K_{b_1b_2b_3b_4}\delta^{10}(k_1+k_2+k_3+k_4)\labell{kin2}
\eeqa
 corresponds to the coupling \reef{Y1} \cite{Gross:1986iv}.   The explicit form of the above amplitude has too many terms to write them all. It has almost all structures of the contractions of the four polarization tensors and the eight momenta. Let us mention which structures the amplitude \reef{kin2}  does not have. The structure of \reef{kin1} dictates that $\Delta{\cal A}$ does not have $(k\inn\veps\inn k)^4$, $k\inn k (k\inn\veps\inn k)^2k\inn\veps\inn\veps \inn k$ and $(k\inn k)^3\,k\inn\veps\inn k\Tr[\veps\inn\veps\inn\veps]$ structures. Obviously, it does not have structures which contain $\Tr[\veps]$ either.  
 
 The couplings \reef{Y2} have been found in \cite{Grisaru:1986vi,Freeman:1986zh} by writing the eight-dimensional tensor $t^{i_1\cdots i_8}t_{j_1\cdots j_8}$ in terms of ten-dimensional tensors. These couplings can also be verified by explicit comparison with the amplitude \reef{kin2}. To this end, one has to calculate the four-graviton amplitude from \reef{Y2} which is 
\beqa
A(1,2,3,4) &=& \gamma \kappa^2e^{-2\phi_0} \bigg[(R_1)_{hmnk}(R_2)_p{}^{mn}{}_q(R_3)^{hrsp}(R_4)^q{}_{rs}{}^k\nonumber\\
&&\qquad\qquad\quad+\frac{1}{2}(R_1)_{hkmn}(R_2)_{pq}{}^{mn}(R_3)^{hrsp}(R_4)^q{}_{rs}{}^k+\cdots\bigg] \labell{onshell}
\eeqa
where dots represent the 23 other   permutation terms. In above amplitude the subscripts $1,2,3,4$ are the particle labels, and 
\beqa
(R_1)^{hmnk}&=&- (\veps_1^{hk}k_1^mk_1^n+\veps_1^{mn}k_1^hk_1^k-\veps_1^{hn}k_1^mk_1^k-\veps_1^{mk}k_1^hk_1^n)
\eeqa
Using the on-shell conditions to write the field theory amplitude in terms of $k_1,k_2,k_3$ and write $k_1\inn\veps_4$ in terms of $k_2\inn\veps_4$ and $k_3\inn\veps_4$, as in string theory amplitude \reef{amp1},  we have found exact agreement with the string theory amplitude   when the polarization tensors are symmetric. 

\section{$(\prt H)^4$ couplings}

The scattering amplitude of four  $B$-fields can be read from \reef{amp1} by using antisymmetric polarizations $\veps^{ab}$. A proposal for the $B$-field couplings in  field theory     is given in \cite{Gross:1986mw} which is the replacement $R_{abcd}\rightarrow R_{abcd}+\kappa e^{-\phi_0/2}H_{ab[c,d]}$. This proposal gives the $(\prt H)^4$ couplings at order $\alpha'^3$ by using the following replacement for the Riemann curvature:
\beqa
R_{abcd}&\rightarrow &\kappa e^{-\phi_0/2}H_{ab[c,d]}\,=\, \kappa e^{-\phi_0/2}(b_{ad,bc}+b_{bc,ad}-b_{ac,bd}-b_{bd,ac})\labell{replace}
\eeqa
We have explicitly check that while the above  replacement in the Lagrangian given in \cite{Gross:1986mw} produces correctly the string amplitude \reef{kin2}, this replacement in the Lagrangian \reef{Y2} does not produce correctly the $B$-field couplings. In particular, the above replacement  in the Lagrangian \reef{Y2} produces terms with structure $k\inn k (k\inn\veps\inn k)^2k\inn\veps\inn\veps \inn k$ whereas the string theory amplitude \reef{kin2} does not produce such structure. The reason for this apparently inconsistency is that the identity that relates the Lagrangian given in \cite{Gross:1986mw} to \reef{Y2} is not an identity any more when one uses the above replacement for the Riemann curvatures in that identity. For example, the Bianchi identity for the   curvature is not an identity when one uses the replacement \reef{replace}.   In this paper, we would like to  find the $B$-field couplings corresponding to the couplings \reef{Y2} by using   the compatibility of this  Lagrangian with the linear T-duality transformations. 

The S-matrix elements in string theory  must satisfy the Ward identity corresponding to the T-duality \cite{ Garousi:2011we,Velni:2012sv}. This means the scattering amplitude \reef{amp1} must be invariant under linear T-duality transformations \reef{linear} on the quantum fluctuations and must be invariant under non-linear T-duality transformation \reef{nonlinear} on the background fields. One can easily verify that the background factor $e^{-2\phi_0}\delta^{10}(k_1+k_2+k_3+k_4)=(2\pi \rho)e^{-2\phi_0}\delta^{9}(k_1+k_2+k_3+k_4)$ where $\rho$ is the radius of the circle along which the T-duality is implemented, is invariant under the T-duality. The amplitude  \reef{kin2} which is the string amplitude  at order $\alpha'^3$, has no massless pole so the T-dual Ward identity dictates that couplings in the spacetime must be invariant under the  linear T-duality.  In spacetime, the invariance of the background under the nonlinear T-duality appears as the invariant of the factor $e^{-2\phi_0}\sqrt{-G}$ in the action.

Now let us apply the linear T-duality   on the quantum fluctuations in  \reef{Y2} to find the couplings of four B-fields. We first use the dimensional reduction to reduce the action to 9 dimensions, and then apply the T-duality transformations on them.  The terms in which the Riemann tensors carry one Killing index $y$ are the following\footnote{From now on we use only subscripts indices and the repeated indices are contracted with the flat metric.}:
\beqa
&&\frac{\gamma e^{-2\phi_0}}{2\kappa^2} \bigg[-8 R_{k n h y} R_{n q p y} R_{k s q y} R_{p s h y}+4 R_{h k n y} R_{p q n y} R_{k s q y} R_{p s h y}\nonumber\\
&&-2 R_{m n k y} R_{m n p y} R_{k s r y} R_{p s r y}-4 R_{k n m y} R_{m p n y} R_{k s r y} R_{p s r y}\bigg]\labell{R41}
\eeqa
We have to find new couplings of four $H$ such that their dimensional reduction transform to the above couplings under the linear T-duality \reef{TR}. Consider the following couplings:
\beqa
&&\frac{\gamma \kappa^2e^{-2\phi_0}}{2} \bigg[-2 H_{h p r,k} H_{n p r,q} H_{k m s,q} H_{m n s,h}-2 H_{h m n,k} H_{h n p,q} H_{k p s,r} H_{m q s,r}\nonumber\\
&&+2 H_{k m n,h} H_{h p q,n} H_{k p s,r} H_{m q s,r}\bigg]\labell{H4}
\eeqa
The dimensional reduction of these couplings produces the following terms: 
\beqa
&&\frac{\gamma \kappa^2e^{-2\phi_0}}{2} \bigg[-8 H_{k m y,q} H_{m n y,h} H_{h p y,k} H_{n p y,q}+4 H_{h m y,k} H_{h p y,q} H_{k p y,r} H_{m q y,r}\nonumber\\
&&-4 H_{m n y,h} H_{h p y,n} H_{m s y,r} H_{p s y,r}-2 H_{h n y,k} H_{h n y,q} H_{k s y,r} H_{q s y,r}\bigg]
\eeqa
which are the transformation of \reef{R41} under the T-duality transformation \reef{TR}. Therefore, the couplings \reef{H4} are the prediction of T-duality for the couplings of four $ \prt H$ at order $\alpha'^3$. We have also calculated its scattering amplitude and find exact agreement with the string theory amplitude \reef{kin2} when the polarization tensors are antisymmetric. Since both the above couplings and the $(\prt H)^4$ couplings in \cite{Gross:1986mw} are reproduced by the string theory amplitude \reef{kin2}, they must be identical up to   some identities.  Extending the linearized couplings \reef{H4} to nonlinear, one finds the couplings in \reef{H40}.

\section{ $R^2(\prt H)^2$  couplings}

There has been one consistency condition for the couplings \reef{H4}, \ie under the dimensional reduction its $H_yH_yH_yH_y$ terms  must be transformed to \reef{R41} under T-duality \reef{TR}. So it was relatively easy to find these terms. The dimensional reduction of the couplings  $RRHH$ however must satisfy four consistency conditions: 1-Their $RRH_yH_y$ terms must transform under T-duality \reef{TR} to the $RRR_yR_y$ terms of the  couplings \reef{Y2}. 2-Their $HHR_yR_y$ terms must transform under T-duality \reef{TR} to  the $HHH_yH_y$ terms of the couplings \reef{H4}. 3-Their $HH_yRR_y$ terms must be invariant under \reef{TR}. 4- Their $H_yH_yR_yR_y$ terms must be  invariant. So it is nontrivial to find such  couplings. 

Let us consider  the $RRR_yR_y$ terms of the dimensional reduction of the couplings \reef{Y2} which are given by 
\beqa
&&\frac{\gamma e^{-2\phi_0}}{2\kappa^2} \bigg[-4 R_{ k n h y } R_{ n q p y } R_{ h r p s } R_{ q r k s }+2 R_{ h k n y } R_{ p q n y } R_{ h r p s } R_{ q r k s }+2 R_{ h k m n } R_{ m n p q } R_{ k s q y } R_{ p s h y }\nonumber\\
&&+4 R_{ h m k n } R_{ m p n q } R_{ k s q y } R_{ p s h y }+4 R_{ m n p q } R_{ m n k y } R_{ q r k s } R_{ p s r y }+8 R_{ k n m y } R_{ m p n q } R_{ q r k s } R_{ p s r y }\bigg]\labell{R42}
\eeqa
One may use the T-duality transformation \reef{TR} to find $RRH_yH_y$ terms and then extend the $y$-index in them to a complete index. In this way one can find the $RRHH$ couplings which are consistent with the above couplings. However, it turns out that their $HH_yRR_y$ terms would not be  invariant under T-duality. They would not be  consistent with the S-matrix element \reef{kin2} either. That means there must be some other terms as well as those  found by using the transformation \reef{TR}.   

The only possibility for extending the transformation \reef{TR} is to add the trivial term $H_{abc,y}$,  \ie   
\beqa
R_{abc y}\rightarrow   -\kappa H_{ab y,c}+\alpha  H_{abc,y}\labell{TR4}
\eeqa
where the coefficient $\alpha$ is an arbitrary  constant. The above extra term in the transformation of the Riemann tensor is zero because of the implicit assumption in the T-duality transformations that fields are independent of the Killing coordinate. However, in extending the $y$-index to a complete index that term makes nontrivial contribution. So we use the above transformation for the Riemann tensor in the couplings \reef{R42} and then extend the $y$-index to a complete index. In doing this one has to use different constants for the coefficients $\alpha$'s in each replacement because a priori we do not know which replacement has such extra term\footnote{The replacement \reef{replace} corresponds to $\alpha=1$. However, as we mentioned in the previous section such replacement in the Lagrangian \reef{Y2} does not produce correctly the $B$-field couplings. }. We have tried to find such coefficients by imposing T-duality transformations. We could not find a consistent set of  coefficients in this way unless we use the on-shell relations. Alternatively, one may  find these constants by comparing the result with the on-shell S-matrix element \reef{kin2}. We find the following result:
\beqa
&&\frac{\gamma  e^{-2\phi_0}}{2}\bigg[4 R_{ h mk n } R_{ m pn q } H_{k r s,q} H_{p r s,h}-8 R_{ h rp s } R_{ q rk s } H_{h k n,m} H_{n p q,m}\nonumber\\
&&+8 R_{ m pn q } R_{ q rk s } H_{h k n,m} H_{h p s,r}+4 R_{ m np q } R_{ q rk s } H_{h m n,k} H_{h p s,r}+8 R_{ m pn q } R_{ q rk s } H_{k m n,h} H_{p r s,h}\nonumber\\
&&+4 R_{ h mk n } R_{ m pn q } H_{h p s,r} H_{k q s,r}-8 R_{ m pn q } R_{ q rk s } H_{k m n,h} H_{h p s,r}\nonumber\\
&&+8 R_{ h mk n } R_{ m pn q } H_{k q s,r} H_{p r s,h}-24 R_{ h rp s } R_{ q rk s } H_{h k n,m} H_{m n q,p}\bigg]\labell{rrhh}
\eeqa
Plus some other terms that their coefficients can not be fixed by the four-point function calculations. They, however, cancels each other when we write them in terms of $h$ and $b$ instead of their field strengths. That means there are identities at four NS-NS level that cancel the terms that are not fixed by the S-matrix calculations. 
Those identities may not hold at five NS-NS level. As a result, there may be some other couplings that their coefficients can be fixed by analyzing the five-point functions.  We have checked that the   couplings \reef{rrhh} satisfy the  above four T-duality constraints.   In checking these constraints,  one has to use the on-shell conditions. Since both the above couplings and the $R^2(\prt H)^2$ couplings in \cite{Gross:1986mw} are produced by the string theory amplitude \reef{kin2}, they must be identical up to some identities. The nonlinear extension of the couplings \reef{rrhh} appears in \reef{rrhh0}.

\section{Dilaton couplings}

The string frame couplings \reef{Y2} and \reef{rrhh0}  can produce various dilaton couplings when transforming them to the Einstein frame. One may then expect that the  dilaton  S-matrix elements   at order $\alpha'^3$ are reproduced by these couplings in the Einstein frame. In this  section we are going to  show that the dilaton couplings in the Einstein frame do not fully reproduce the string theory amplitudes. Hence, the string frame field theory should contain some new dilaton couplings. 

The S-matrix element of one dilaton and three gravitons in string theory side is given by \reef{kin2} in which one of the polarizations is \reef{del} and the other three are symmetric and traceless. On the other hand, the scattering amplitude of four symmetric polarization tensors must satisfy the Ward identity, that is, if one replaces each polarization tensor by $\veps^{ab}\rightarrow k^a\z^b+k^b\z^a$ where    $\z^a$ is an arbitrary vector, the amplitude must be  zero. This indicates that the term $-k^a\ell^b-k^b\ell^a$ in the dilaton polarization \reef{del} must disappear in the string amplitude of one dilaton and three symmetric polarization tensors, \ie the dilaton polarization is effectively $\veps_1^{ab}=\phi_1\eta^{ab}/\sqrt{8}$. This replacement cancels many terms in \reef{kin2}. The surviving terms are the following: 
\beqa
\Delta{\cal A} &=& \frac{\gamma \kappa^2e^{-2\phi_0}}{2\sqrt{8} }\bigg[ 16 \left(k_2.k_3\right){}^2 \left(k_3.k_1\right){}^2 \Tr\left[ \veps _2\right] \Tr\left[\veps _3.\veps _4\right]+ 16 \left(k_2.k_1\right){}^2 k_1.\veps _3.k_1 k_2.\veps _4.k_2 \Tr\left[\veps _2\right]\nonumber\\
&&+32 k_2.k_1 k_3.k_1 k_1.\veps _3.k_1 k_2.\veps _4.k_2 \Tr\left[\veps _2\right]+16 \left(k_3.k_1\right){}^2 k_1.\veps _3.k_1 k_2.\veps _4.k_2 \Tr\left[\veps _2\right]\nonumber\\
&&+32 k_2.k_1 k_3.k_1 k_1.\veps _3.k_2 k_2.\veps _4.k_2 \Tr\left[\veps _2\right]+32 \left(k_3.k_1\right){}^2 k_1.\veps _3.k_2 k_2.\veps _4.k_2 \Tr\left[\veps _2\right]\nonumber\\
&&+16 \left(k_3.k_1\right){}^2 k_2.\veps _3.k_2 k_2.\veps _4.k_2 \Tr\left[\veps _2\right]+32 k_2.k_1 k_3.k_1 k_1.\veps _3.k_2 k_2.\veps _4.k_3 \Tr\left[\veps _2\right]\nonumber\\
&&+32 \left(k_3.k_1\right){}^2 k_1.\veps _3.k_2 k_2.\veps _4.k_3 \Tr\left[\veps _2\right]+32 \left(k_3.k_1\right){}^2 k_2.\veps _3.k_2 k_2.\veps _4.k_3 \Tr\left[\veps _2\right]\nonumber\\
&&+16 \left(k_3.k_1\right){}^2 k_2.\veps _3.k_2 k_3.\veps _4.k_3 \Tr\left[\veps _2\right]+32 \left(k_2.k_1\right){}^2 k_3.k_1 k_1.\veps _3.\veps _4.k_2 \Tr\left[\veps _2\right]\nonumber\\
&&+64 k_2.k_1 \left(k_3.k_1\right){}^2 k_1.\veps _3.\veps _4.k_2 \Tr\left[\veps _2\right]+32 \left(k_3.k_1\right){}^3 k_1.\veps _3.\veps _4.k_2 \Tr\left[\veps _2\right]\nonumber\\
&&+32 k_2.k_1 \left(k_3.k_1\right){}^2 k_2.\veps _3.\veps _4.k_2 \Tr\left[\veps _2\right]+32 \left(k_3.k_1\right){}^3 k_2.\veps _3.\veps _4.k_2 \Tr\left[\veps _2\right]\nonumber\\
&&+32 k_2.k_1 \left(k_3.k_1\right){}^2 k_2.\veps _3.\veps _4.k_3 \Tr\left[\veps _2\right]+32 \left(k_3.k_1\right){}^3 k_2.\veps _3.\veps _4.k_3 \Tr\left[\veps _2\right]\bigg]\phi_1\nonumber\\
&&+(2\leftrightarrow 3)+(2\leftrightarrow 4) \labell{phi1}
\eeqa
which are zero when the polarizations are traceless. Note that the  terms in \reef{kin2} which contain the trace of four polarization tensors, \eg $\Tr[\veps_1\inn\veps_2\inn\veps_3\inn\veps_4]$,  are canceled when one of the polarization is replace by $\phi\eta_{ab}/\sqrt{8}$.   

The string scattering amplitude produces couplings in the Einstein frame, so in the field theory side  we consider the transformation of the string couplings \reef{Y2} to the Einstein frame, \ie, $G_{ab}=e^{\Phi/2}G^E_{ab}$. At the linear order it gives $h_{ab}=h^E_{ab}+\phi\eta_{ab}\sqrt{8}$, and in terms of the linearized Riemann curvature it becomes $R_{ab}{}^{cd}=R^E_{ab}{}^{cd}-\kappa\eta^{[c}_a\phi_{,b]}{}^{d]}/\sqrt{8}$.   In the field theory couplings \reef{Y2} 
,    one must then replace one of the polarizations by $\phi\eta_{ab}/\sqrt{8}$, hence, one  again finds zero result for the scattering amplitude of one dilaton and three gravitons. So it confirms that there is no coupling of one dilaton and three gravitons in  the string frame  \cite{Candelas:1986tz,Myers:1987qx} or in the Einstein frame. This is not the case, however,  for the couplings of two dilatons and two gravitons as we shall see below.

The scattering amplitude of two dilatons and two symmetric tensors in string theory side can be read from the  amplitude \reef{phi1} by replacing the polarization $\veps_2$ with \reef{del}. Apart from the terms containing the trace of $\veps_2$ which is $\Tr[\veps_2]=\phi_2\sqrt{8}$, the auxiliary term $-k_2^a\ell_2^b-k_2^b\ell_2^a$ in the dilaton polarization \reef{del}  cancels in the terms in the last line of \reef{phi1}, hence, effectively  for these terms the dilaton polarization is $\phi_2\eta_{ab}/\sqrt{8}$.  
The amplitude becomes
\beqa
\Delta{\cal A}  &=& \frac{\gamma \kappa^2e^{-2\phi_0}}{ 2}\bigg[ 16 \left(k_2.k_3\right){}^2 \left(k_3.k_1\right){}^2   \Tr\left[\veps _3.\veps _4\right]+ 16 \left(k_2.k_1\right){}^2 k_1.\veps _3.k_1 k_2.\veps _4.k_2  \nonumber\\
&&+32 k_2.k_1 k_3.k_1 k_1.\veps _3.k_1 k_2.\veps _4.k_2  +16 \left(k_3.k_1\right){}^2 k_1.\veps _3.k_1 k_2.\veps _4.k_2  \nonumber\\
&&+32 k_2.k_1 k_3.k_1 k_1.\veps _3.k_2 k_2.\veps _4.k_2  +32 \left(k_3.k_1\right){}^2 k_1.\veps _3.k_2 k_2.\veps _4.k_2  \nonumber\\
&&+16 \left(k_3.k_1\right){}^2 k_2.\veps _3.k_2 k_2.\veps _4.k_2  +32 k_2.k_1 k_3.k_1 k_1.\veps _3.k_2 k_2.\veps _4.k_3  \nonumber\\
&&+32 \left(k_3.k_1\right){}^2 k_1.\veps _3.k_2 k_2.\veps _4.k_3 +32 \left(k_3.k_1\right){}^2 k_2.\veps _3.k_2 k_2.\veps _4.k_3  \nonumber\\
&&+16 \left(k_3.k_1\right){}^2 k_2.\veps _3.k_2 k_3.\veps _4.k_3  +32 \left(k_2.k_1\right){}^2 k_3.k_1 k_1.\veps _3.\veps _4.k_2  \nonumber\\
&&+64 k_2.k_1 \left(k_3.k_1\right){}^2 k_1.\veps _3.\veps _4.k_2  +32 \left(k_3.k_1\right){}^3 k_1.\veps _3.\veps _4.k_2  \nonumber\\
&&+32 k_2.k_1 \left(k_3.k_1\right){}^2 k_2.\veps _3.\veps _4.k_2  +32 \left(k_3.k_1\right){}^3 k_2.\veps _3.\veps _4.k_2  \nonumber\\
&&+32 k_2.k_1 \left(k_3.k_1\right){}^2 k_2.\veps _3.\veps _4.k_3  +32 \left(k_3.k_1\right){}^3 k_2.\veps _3.\veps _4.k_3 \nonumber\\
&&+2 \left(k_2.k_1\right){}^2 \left(\left(k_2.k_1\right){}^2+2 k_2.k_1 k_3.k_1+2 \left(k_3.k_1\right){}^2\right)  {\Tr}\left[\veps _3\right]  {\Tr}\left[\veps _4\right]\bigg]\phi_1\phi_2\labell{amp2}  
\eeqa
where the terms in the last line of \reef{phi1} appear in the last line of the above amplitude. Apart from these terms   which are zero for graviton, all other terms are  in fact the terms of the scattering amplitude \reef{kin2} which include\footnote{Note that the proposal given in \cite{Gross:1986mw} that extends  the Riemann curvature to include the dilaton, \ie $R_{ab}{}^{cd}\rightarrow R_{ab}{}^{cd}-\kappa\eta^{[c}_a\phi_{,b]}{}^{d]}/\sqrt{8}$, is equivalent to extension $h_{ab}\rightarrow h_{ab}+\phi\eta_{ab}/\sqrt{8}$. This  gives $\Tr[\veps_1\inn\veps_2]\rightarrow \Tr[\veps_1\inn\veps_2]+\phi_1\phi_2$  in  the eight-dimensional transverse space of the light-cone formalism.} $\Tr[\veps_1\inn\veps_2]=\phi_1\phi_2$. 

In transforming the couplings \reef{Y2} to the Einstein frame, one transforms $h_{ab}h^{ba}= h^E_{ab}(h^E)^{ba}+\frac{10}{8}\phi^2$, or in terms of polarization it becomes $\Tr[\veps_1\inn\veps_2]=\Tr[\veps^E_1\inn\veps^E_2]+\frac{10}{8}\phi_1\phi_2$. So the above amplitude is not fully reproduced by transforming the string frame couplings \reef{Y2} to the Einstein frame, \ie $5/4$ of the above amplitude is reproduced by \reef{Y2} and $-1/4$ of it is a new dilaton coupling in the string frame. 

To find the field theory couplings corresponding to the above amplitude, we have to find the couplings in \reef{Y2} which have $h_{ab}h^{ba}$ and use the replacement $h_{ab}h^{ba}\rightarrow  \phi^2$ in them. On the other hand, in the dimensional reduction the term $h_{yy}h^{yy}$ is a component of $h_{ab}h^{ba}$. Hence, to find the couplings corresponding to the above amplitude, we have to find the $h_{yy}h^{yy}$-terms in the dimensional reduction of  \reef{Y2} and use the replacement $h_{yy}h^{yy}\rightarrow  \phi^2$ in them. The dimensional reduction produces the following terms: 
\beqa
\frac{\gamma e^{-2\phi_0}}{2\kappa^2} \bigg[R_{h k m n} R_{m n p q} R_{h y p y} R_{k y q y}+2 R_{h m k n} R_{m p n q} R_{h y p y} R_{k y q y}+2 R_{h r p s} R_{q r k s} R_{h y k y} R_{p y q y}\bigg]\labell{Ryy}
\eeqa
Using the fact that $R_{h y p y}=-\frac{\rho^2\kappa}{\alpha'} h_{yy,hp}$, one finds    the couplings corresponding to the amplitude \reef{amp2} to be
\beqa
\frac{\gamma  e^{-2\phi_0}}{2} \bigg[R_{h k m n} R_{m n p q} \phi_{,hp} \phi_{,kq}  +2 R_{h m k n} R_{m p n q} \phi_{,hp}\phi_{,kq} +2 R_{h r p s} R_{q r k s} \phi_{,hk}\phi_{,pq} \bigg]\labell{delRR}
\eeqa
We have also checked it  explicitly that the above couplings produce the amplitude \reef{amp2}. These couplings are also invariant under linear T-duality because the T-dual extension of the second derivative of the dilaton \reef{TR2} contains the Ricci tensor  which is zero on-shell. The nonlinear extension of the above  couplings with the factor of $-1/4$ appears in \reef{delRR0}.

The scattering amplitude of three dilatons and one symmetric tensor  is given by the scattering amplitude of two dilatons and two symmetric tensors \reef{amp2} in which one of the symmetric tensor is \reef{del}. For the term in the last line one must replace $\Tr\left[\veps _3\right]= \phi_3\sqrt{8}$ and for all other terms one must replace $(\veps_3)_{ab}=\phi_3\eta_{ab}/\sqrt{8}$. The result is  
\beqa
\Delta{\cal A} &=& 2\sqrt{2} \gamma \kappa^2e^{-2\phi_0}\bigg[  \left(\left(k_2.k_1\right){}^2+k_2.k_1 k_3.k_1+\left(k_3.k_1\right){}^2\right){}^2  
{\Tr}\left[\veps _4\right]\bigg] \phi_1\phi_2\phi_3  
\eeqa
 which is zero when the polarization tensor is traceless. Hence, there is no coupling of three dilatons and one graviton.     

The scattering amplitude of four dilatons is given by the above   amplitude in which the trace of $\veps_4$ is replace by $\Tr[\veps_4]=\phi_4\sqrt{8}$, \ie  
\beqa
\Delta{\cal A} &=& 8 \gamma \kappa^2e^{-2\phi_0}\bigg[  \left(\left(k_2.k_1\right){}^2+k_2.k_1 k_3.k_1+\left(k_3.k_1\right){}^2\right){}^2  
\bigg] \phi_1\phi_2\phi_3\phi_4  \labell{extra2}
\eeqa
The above terms are in fact the  terms of the scattering amplitude of four symmetric tensors \reef{kin2}  which include the trace of two polarization tensors, \eg $\Tr[\veps_1\inn\veps_2]\Tr[\veps_3\inn\veps_4]=\phi_1\phi_2\phi_3\phi_4$. On the other hand, the $R_{yy}R_{yy}R_{yy}R_{yy}$ terms of the dimensional reduction of \reef{Y2} produce the trace of two polarization tensors. In fact the traces of four polarization tensors \eg $\Tr[\veps_1\inn\veps_2\inn\veps_3\inn\veps_4]$ in the amplitude \reef{kin2} are canceled when the polarizations commute inside the trace which is the case for the component $\veps_{yy}$ which appears in the couplings $R_{yy}R_{yy}R_{yy}R_{yy}$. So the couplings corresponding to the above amplitude  can be read from  $R_{yy}R_{yy}R_{yy}R_{yy}$  which are
\beqa
\frac{\gamma e^{-2\phi_0}}{ 2\kappa^2}\bigg[-2 R_{ h y n y } R_{ h y s y } R_{ n y q y } R_{ q y s y }+2 R_{ m y n y }R_{ m y n y } R_{ r y s y } R_{ r y s y }\bigg] 
\eeqa
Inspired by these couplings, one finds  the couplings corresponding to \reef{extra2} to be 
\beqa
\frac{\gamma \kappa^2e^{-2\phi_0}}{ 2}\bigg[-2 \phi_{,hn}  \phi_{,hs} \phi_{,nq}  \phi_{,qs} +2 \phi_{,mn} \phi_{,mn} \phi_{,rs} \phi_{,rs}  \bigg]\labell{phi2}
\eeqa
We have also checked the above couplings  by   direct comparison with the amplitude \reef{extra2}.   In transforming the couplings \reef{Y2} and \reef{delRR0} to the Einstein frame one transforms $\Tr[\veps_i\inn\veps_j]\Tr[\veps_k\inn\veps_l]$ in \reef{Y2} to $(\frac{10}{8})^2\phi_i\phi_j\phi_k\phi_l$, and $-\frac{1}{4}\phi_1\phi_2\Tr[\veps_3\inn\veps_4]$ in \reef{delRR0} to $-\frac{5}{8}\phi_1\phi_2\phi_3\phi_4$. So $25/16-10/16$ of the above amplitude is reproduced by transforming the couplings \reef{Y2} and \reef{delRR0} to the Einstein frame and $1/16$ of it is a new dilaton coupling in the string frame. The  nonlinear extension  of \reef{phi2} with the factor of $1/16$ appears in \reef{del4}.

We finally consider the couplings involving the dilaton and the $B$-field. The scattering amplitude of two symmetric tensors and two B-fields has no trace of one symmetric tensor, consequently, the scattering amplitude of one dilaton, one symmetric tensor  and two B-fields   is given by the former  amplitude   in which one of the symmetric tensor is replaced by $\phi\eta_{ab}/\sqrt{8}$. The result is 
\beqa
\Delta{\cal A}&=&\frac{\gamma \kappa^2e^{-2\phi_0}}{ 2\sqrt{8}}\bigg[32 k_2.k_1 k_3.k_1 k_1.\veps _3.k_2 k_2.\veps _4.k_3+32 \left(k_3.k_1\right){}^2 k_1.\veps _3.k_2 k_2.\veps _4.k_3\nonumber\\&&-32 \left(k_2.k_1\right){}^2 k_3.k_1 k_1.\veps _3.\veps _4.k_2-64 k_2.k_1 \left(k_3.k_1\right){}^2 k_1.\veps _3.\veps _4.k_2\nonumber\\&&-32 \left(k_3.k_1\right){}^3 k_1.\veps _3.\veps _4.k_2-32 k_2.k_1 \left(k_3.k_1\right){}^2 k_2.\veps _3.\veps _4.k_2\nonumber\\&&-32 \left(k_3.k_1\right){}^3 k_2.\veps _3.\veps _4.k_2-32 k_2.k_1 \left(k_3.k_1\right){}^2 k_2.\veps _3.\veps _4.k_3\nonumber\\&&-32 \left(k_3.k_1\right){}^3 k_2.\veps _3.\veps _4.k_3-16 \left(k_2.k_1\right){}^2 \left(k_3.k_1\right){}^2 \Tr\left[\veps _3.\veps _4\right]\nonumber\\&&-32 k_2.k_1 \left(k_3.k_1\right){}^3 \Tr\left[\veps _3.\veps _4\right]-16 \left(k_3.k_1\right){}^4 \Tr\left[\veps _3.\veps _4\right]\bigg]\Tr[\veps_2]\phi_1+\cdots \labell{extra3}
\eeqa
where dots refer to the terms which are not proportional to $\Tr[\veps_2]$. They are 
reproduced by transforming the couplings \reef{rrhh} to the  Einstein frame. So there is no coupling of one dilaton, one graviton and two $B$-fields in the string frame. 

The scattering amplitude of two dilatons   and two B-fields is given by the  amplitude \reef{extra3} in which the symmetric polarization is \reef{del}. The terms in which the   polarization appears as   $\Tr[\veps_2] $, are invariant under the Ward identity associated with the symmetric tensor. For these terms one should replace $\Tr[\veps_2]=\phi_2\sqrt{8} $. The result is  the following:
\beqa
\Delta{\cal A}_1&=&\frac{\gamma \kappa^2e^{-2\phi_0}}{ 2 }\bigg[16 k_3.k_1 k_3.k_2 \left(-2 k_1.\epsilon _3.k_2 k_2.\epsilon _4.k_3+k_2.k_1 \left(2 k_1.\epsilon _3.\epsilon _4.k_2+k_3.k_1  {\Tr}\left[\epsilon _3.\epsilon _4\right]\right)\right.\nonumber\\
&&\left.+k_3.k_1 \left(2 \left(k_1.\epsilon _3.\epsilon _4.k_2+k_2.\epsilon _3.\epsilon _4.k_2+k_2.\epsilon _3.\epsilon _4.k_3\right)+k_3.k_1  {\Tr}\left[\epsilon _3.\epsilon _4\right]\right)\right)\bigg] \phi_1\phi_2  \labell{extra4}
\eeqa
The other terms separately satisfy the Ward identity associated with the symmetric tensor. So   $-k_2^a\ell_2^b-k_2^b\ell_2^a$ in the dilaton polarization \reef{del}  cancels in these terms, hence, effectively    the dilaton polarization is $\phi_2\eta_{ab}/\sqrt{8}$. The result in this case is
\beqa
\Delta{\cal A}_2&=&\frac{\gamma \kappa^2e^{-2\phi_0}}{ 2 }\bigg[-8( k_2.k_1 )^2 \left(-2 k_1.\epsilon _3.k_2 k_2.\epsilon _4.k_3+k_2.k_1 \left(2 k_1.\epsilon _3.\epsilon _4.k_2+k_3.k_1  {\Tr}\left[\epsilon _3.\epsilon _4\right]\right)\right.\nonumber\\
&&\left.+k_3.k_1 \left(2 \left(k_1.\epsilon _3.\epsilon _4.k_2+k_2.\epsilon _3.\epsilon _4.k_2+k_2.\epsilon _3.\epsilon _4.k_3\right)+k_3.k_1  {\Tr}\left[\epsilon _3.\epsilon _4\right]\right)\right)\bigg] \phi_1\phi_2  \labell{extra5}
\eeqa
The above amplitude is reproduced by transforming the couplings \reef{rrhh} to the Einstein frame. 

To find the couplings corresponding to the  amplitude \reef{extra4}, we note that these terms are the terms of the scattering amplitude of two symmetric tensors and two B-fields which are proportional to $\Tr[\veps_1\inn\veps_2]=\phi_1\phi_2$. So the couplings corresponding to \reef{extra4} may be read from the   $HHR_{yy}R_{yy}$ terms of the dimensional reduction of the couplings \reef{rrhh} which are
\beqa
&&\frac{\gamma  e^{-2\phi_0}}{ 2}\bigg[-24 R_{ h{  }y p y } R_{ k y q y } H_{h{  }k n,m} H_{m n q,p}-8 R_{ h{  }y p y } R_{ k y q y } H_{h{  }k n,m} H_{n p q,m}\nonumber\\&&+4 R_{ h{  }y k y } R_{ p y q y } H_{h{  }p s,r} H_{k q s,r}+8 R_{ h{  }y k y } R_{ p y q y } H_{k q s,r} H_{p r s,h }+4 R_{ h{  }y k y } R_{ p y q y } H_{k r s,q} H_{p r s,h }\bigg]\nonumber
\eeqa
Inspired by this, one finds  the following couplings of two dilatons and two $B$-fields:
\beqa
&&\frac{\gamma \kappa^2e^{-2\phi_0}}{ 2}\bigg[-24 \phi_{,hp}  \phi_{,kq} H_{h{  }k n,m} H_{m n q,p}-8\phi_{,hp}   \phi_{,kq}  H_{h{  }k n,m} H_{n p q,m}\nonumber\\&&+4 \phi_{,hk}  \phi_{,pq} H_{h{  }p s,r} H_{k q s,r}+8 \phi_{,hk}  \phi_{,pq}  H_{k q s,r} H_{p r s,h } +4 \phi_{,hk}  \phi_{,pq}  H_{k r s,q} H_{p r s,h }\bigg]\nonumber
\eeqa
We have checked explicitly that the above couplings produce the amplitude \reef{extra3}. Here again $5/4$ of the above couplings are reproduced by transforming the couplings \reef{rrhh} to the Einstein frame, and $-1/4$ of them are new couplings. The nonlinear extension of these   couplings appear in \reef{phiH}.

\section{Discussion}

In this paper we have extended the sigma model  Riemann curvature couplings \reef{Y2} to include the $B$-field and the dilaton couplings. We have found these new couplings by imposing the consistency of the couplings \reef{Y2} with the linear T-duality and by the S-matrix calculations. The T-duality  in these couplings   is satisfied on-shell. Even in the absence of the $B$-field, the couplings \reef{Y2} satisfy the standard T-duality only on-shell. The reason is that  the dimensional reduction of the couplings \reef{Y2} contains  the following term:
\beqa
  R_{ k yn y } R^{ n}{}_{ yq y } R_{ r ys y }R^{ q rk s }\eta^{yy}\eta^{yy}\eta^{yy}\labell{non}
\eeqa
which is not  invariant under the T-duality \reef{TR1}. However, using the same calculation as we have done in \reef{onshell} one finds the on-shell amplitude corresponding to this coupling is zero. This may be the reason that the $RRHH$ couplings \reef{rrhh} are also invariant under on-shell T-duality. 

In general one expects the effective actions to be invariant under off-shell T-duality. So the effective action which includes the supergravity at order $\alpha'^0$ and the Riemann curvature corrections \reef{Y2} at order $\alpha'^3$ should  be invariant under an off-shell T-duality which   receives quantum corrections. In fact there are different sets of Riemann curvature corrections which are related to each others via some couplings involving the Ricci and scalar curvatures \cite{Myers:1987qx}. These terms can be eliminated by field redefinitions involving higher derivative terms. The field redefinitions at the same time changes the standard form of the T-duality \reef{nonlinear} to a non-standard form which receives the higher derivative corrections.  So one expects one set of Riemann curvature corrections to be invariant under the standard T-duality trnsformations, and all other sets to be invariant under the non-standard T-duality transformations.  

We have found  four NS-NS couplings which are related to the four-graviton couplings \reef{Y2} by on-shell linear T-duality transformations. However, there are ambiguities in the couplings \reef{Y2} which can be fixed by studying the five-graviton amplitudes. For example, the four Riemann curvature couplings $\eps_{10}\cdot\eps_{10}RRRR$ can be added to \reef{Y2} because this term has its first non-zero contribution at  five gravitons \cite{Zumino:1985dp}. The sigma-model approach implies that this term appears in the effective action  \cite{Grisaru:1986px}. It would be interesting to find the couplings which are related to the couplings $\eps_{10}\cdot\eps_{10}RRRR$ under T-dual Ward identity.

We have found the couplings \reef{H40} and \reef{rrhh0} by using the fact that the S-matrix elements should satisfy the T-dual Ward identity \cite{Garousi:2011we,Velni:2012sv}. On the other hand  the S-matrix elements should satisfy the S-dual Ward identity \cite{Garousi:2011we, Garousi:2011vs,Garousi:2011jh,Garousi:2012gh}. Using this identity, one may extend the couplings we have found in this paper  to include the R-R couplings as well. The couplings involving the R-R two-form can easily be included in  \reef{H40} and \reef{rrhh0}  by replacing $e^{-\phi_0}H_{abc}H_{def}$ with the following S-duality invariant expression:
\beqa
e^{-\phi_0}H_{abc,d}H_{efg,h}\rightarrow e^{-\phi_0}H_{abc,d}H_{efg,h}+ e^{\phi_0}F_{abc,d}F_{efg,h}
  \nonumber
\eeqa
where $F$ is the field strength of the R-R two-form. Similar extension for  the D-brane couplings at order $\alpha'^2$ has been verified by explicit calculations in \cite{Garousi:2011fc}. 
A representation for the  $R^2(\prt F)^2$ couplings have been found in \cite{Peeters:2003pv}. It has been shown in \cite{Peeters:2003pv} that  this representation is the same as the $R^2(\prt F)^2$ couplings that one finds by using the above extension in the  $R^2(\prt H)^2$ couplings in \cite{Gross:1986mw}. The $R^2(\prt F)^2$ terms that we have found are then the same as the couplings found in \cite{Peeters:2003pv} up to some identities.  One may  use the consistency of the above R-R two-form couplings  with the linear T-duality to find all other R-R couplings at order $\alpha'^3$. One may also extend the four-point couplings at order $\alpha'^3$ to arbitrary order of $\alpha'$  using the prescription given in \cite{Chandia:2003sh}. 

{\bf Acknowledgments}:   I would like to thank A. Ghodsi and R. Medina for useful discussions. This work is supported by Ferdowsi University of Mashhad under grant 2/23265-1391/07/18.

\end{document}